\documentclass[12pt,a4paper]{iopart}
\pdfoutput=1 
\usepackage{graphicx}
\usepackage[utf8]{inputenc} 
\usepackage{color}
\usepackage[
,textwidth=15cm
,textheight=20cm
,verbose
,dvips
]{geometry}
\usepackage[numbers,comma,square,sort&compress,merge]{natbib}

\begin{document}

\title[Correlation between the structural and optical properties of GaN nanowires]{Correlation between the structural and optical properties of spontaneously formed GaN nanowires: a quantitative evaluation of the impact of nanowire coalescence}

\author{S Fernández-Garrido, V M Kaganer, C Hauswald, B Jenichen, M Ramsteiner, V Consonni\footnote{Present address: Laboratoire des Matériaux et du Génie Physique, CNRS - Grenoble INP, 3 parvis Louis Néel, 38016 Grenoble, France}, L Geelhaar and O Brandt}
\address{Paul-Drude-Institut für Festkörperelektronik,
Hausvogteiplatz 5--7, 10117 Berlin, Germany}
\ead{garrido@pdi-berlin.de}

\begin{abstract}
We investigate the structural and optical properties of spontaneously formed GaN nanowires with different degrees of coalescence. This quantity is determined by an analysis of the cross-sectional area and perimeter of the nanowires obtained by plan-view scanning electron microscopy. X-ray diffraction experiments are used to measure the inhomogeneous strain in the nanowire ensembles as well as the orientational distribution of the nanowires. The comparison of the results obtained for GaN nanowire ensembles prepared on bare Si$(111)$ and AlN buffered 6H-SiC$(000\bar{1})$ reveals that the main source of the inhomogeneous strain is the random distortions caused by the coalescence of adjacent nanowires. The magnitude of the strain inhomogeneity induced by nanowire coalescence is found not to be determined solely by the coalescence degree, but also  by the mutual misorientation of the coalesced nanowires. The linewidth of the donor-bound exciton transition in photoluminescence spectra does not exhibit a monotonic increase with the coalescence degree. In contrast, the comparison of the root mean square strain with the linewidth of the donor-bound exciton transition reveals a clear correlation: the higher the strain inhomogeneity, the larger the linewidth.
\end{abstract}



\maketitle

\section{Introduction}

Semiconductor nanowires (NWs) lift most of the fundamental constraints in epitaxial growth and provide new degrees of freedom for the design of electronic and optoelectronic devices. In contrast to epitaxial films, single-crystalline semiconductors can be grown in form of NWs on dissimilar substrates because their high aspect ratio inhibits the propagation of dislocations along the NW axis \cite{Tomioka_nl_2008,Urban_njp_2013}. Furthermore, their aspect ratio facilitates an efficient elastic strain relaxation at the NW sidewalls. These virtues enable the fabrication of axial heterostructures of materials that are, in general, incompatible due to large mismatches in lattice constants and thermal expansion coefficients \cite{Bjork_nl_2002,Gudiksen_nature_2002,Johansson_CrystEngComm_2011}. Consequently, NWs have attracted great interest for the fabrication of semiconductor devices on low cost substrates and for the monolithic integration of optoelectronic devices on silicon \cite{Martensson2004,lagally2004,Tomioka_nl_2010,Li_joap_2012}.

In plasma-assisted molecular beam epitaxy (PA-MBE), the technologically relevant compound semiconductor GaN has a pronounced tendency to spontaneously form dense arrays ($10^{9}-10^{10}$~cm$^{-2}$) of NWs on crystalline as well as amorphous substrates under specific growth conditions \cite{Calarco2007,Songmuang_apl_2007,Stoica_2008,Fernandez-Garrido2009,Geelhaar_ieeejstqe_2011,Consonni_2011_b,Bertness_q_2011,Consonni_pssrrl_2013,Sobanska_jap_2014}. Regardless of the substrate used for growth, spontaneously formed GaN NWs are initially single crystalline and free of homogeneous strain \cite{Geelhaar_ieeejstqe_2011,Bertness_q_2011,Calleja2000,Trampert2003,Jenichen2011a}. However, GaN NW ensembles frequently exhibit a high degree of coalescence \cite{Calarco2007,Fernandez-Garrido2009,Consonni_2011_b,Park_nanotech_2006,Grossklaus_jcg_2013,Brandt_unpublished_2013,Fan_jovstb_2014} and inhomogeneous strain on a microscopic scale, i.\,e., micro-strain \cite{Jenichen2011a,Kaganer_prb_2012}.

The origin of the inhomogeneous strain detected in x-ray diffraction (XRD) experiments is not well understood yet. The main sources of inhomogeneous strain suggested in the literature for GaN NW ensembles grown on bare Si$(111)$ are the random distortions at the interface between the NWs and the substrate \cite{Kaganer_prb_2012}, where an amorphous Si$_{x}$N$_{y}$ interlayer is formed due to the unavoidable nitridation of the substrate prior to the nucleation of GaN NWs \cite{Consonni_pssrrl_2013,Trampert2003,Wierzbicka_natech_2012,Hestroffer_apl_2012}, and strain introduced at coalescence boundaries \cite{Jenichen2011a,Kaganer_prb_2012}.

The coalescence of spontaneously formed GaN NWs is the consequence of their spatially random nucleation on the substrate \cite{Consonni_pssrrl_2013,Ristic_jcg_2008}, their radial growth following nucleation \cite{Fan_jovstb_2014,Fernandez-Garrido_nl_2013}, and their mutual misorientation together with their axial growth. \cite{Jenichen2011a,Fan_jovstb_2014,Wierzbicka_natech_2012,Horak_jap_2008}. Consequently, the degree of coalescence depends on the type of substrate as well as on the nucleation and growth conditions, namely, the substrate temperature, the impinging fluxes and the growth time.\cite{Park_nanotech_2006,Fernandez-Garrido2009,Grossklaus_jcg_2013,Consonni_pssrrl_2013,Fan_jovstb_2014} As result of the mutual misorientation of coalesced NWs, networks of dislocations and basal-plane stacking faults (SFs) are formed at the coalescence joints if the tilt and twist of the coalesced NWs cannot be accommodated elastically \cite{Jenichen2011a,Fan_jovstb_2014,Consonni2009}.

Due to the strong impact of coalescence on the properties of GaN NW ensembles, several groups have attempted to correlate the coalescence degree and the structural and optical properties of GaN NWs \cite{Jenichen2011a,Park_nanotech_2006,Lefebvre2011}. However, these studies suffered from the lack of objective methods to distinguish single NWs from coalesced aggregates.

In this work, we employ the methods recently proposed in Ref.\cite{Brandt_unpublished_2013} to quantify the coalescence degree of NW ensembles prepared on different types of substrates. Subsequently, we investigate the optical and structural properties of the samples using a variety of experimental techniques and compare them with the coalescence degree of the NW ensemble. The results evidence that the inhomogeneous strain detected in the NWs is mainly caused by their coalescence. The magnitude of the strain inhomogeneity is found to depend on the coalescence degree as well as on the mutual misorientation of adjacent NWs. Finally, the analysis of the optical properties of GaN NWs as a function of the strain inhomogeneity reveals a clear correlation: the higher the root mean square (rms) strain, the larger the linewidths of the excitonic transitions observed by photoluminescence (PL) spectroscopy. 

\section{Experiment}

\begin{figure}
\hspace{1in}\includegraphics {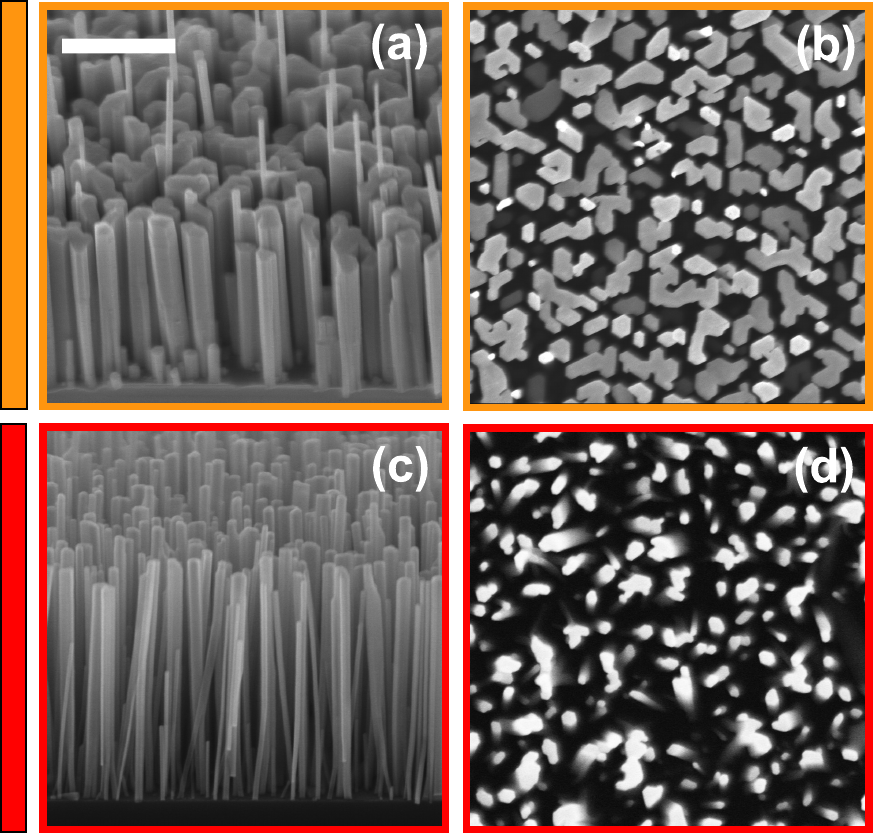} 
\caption{\label{SEM_Si}(Color online) Bird's eye and plan-view SEM micrographs of sample A [(a) and (b)] and sample B [(c) and (d)]. The magnification is the same for all images and the scale bar shown in (a) represents 500 nm. The color bars on the left side indicate the color code used throughout the manuscript.}
\end{figure}

\begin{figure}
\hspace{1in}\includegraphics {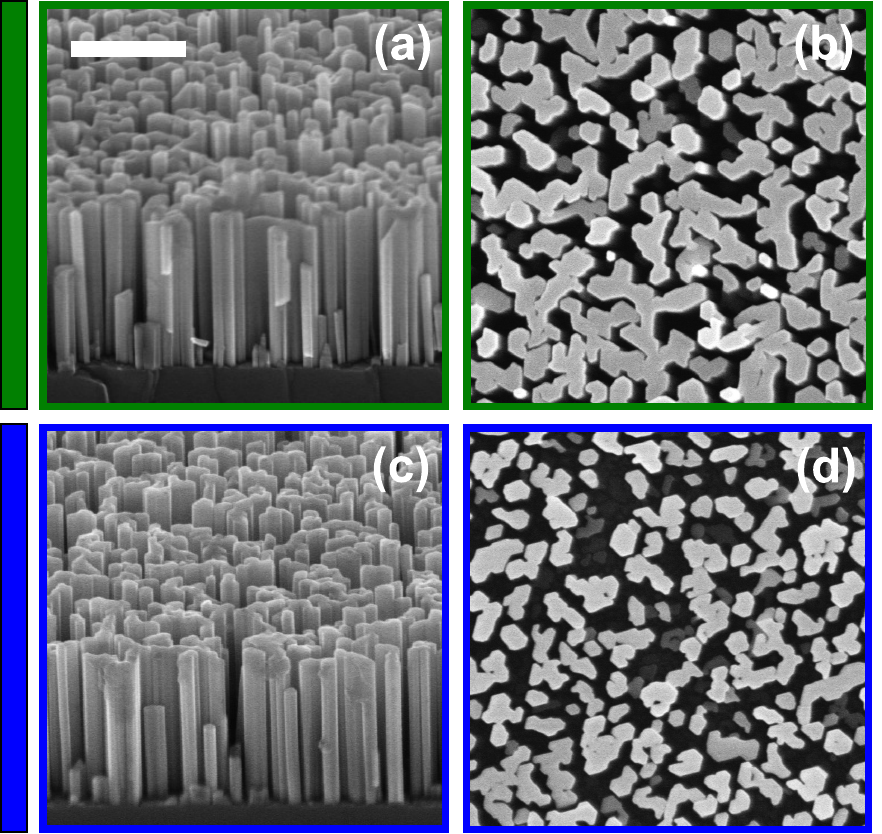}
\caption{\label{SEM_SiC}(Color online) Bird's eye and plan-view SEM micrographs of 
sample C [(a) and (b)] and sample D [(c) and (d)]. The magnification is the same for all images and the scale bar shown in (a) represents 500 nm. The color bars on the left side indicate the color code used throughout the manuscript.}
\end{figure}

The GaN NW ensembles studied here formed spontaneously during PA-MBE \cite{Geelhaar_ieeejstqe_2011,Bertness_q_2011,Consonni_pssrrl_2013} on substrates of two different types: bare Si$(111)$ and AlN-buffered 6H-SiC$(000\overline{1})$. For each substrate, appropriate growth conditions were chosen for the synthesis of two GaN NW samples with different degrees of coalescence following the methods reported in Refs.\cite{Fernandez-Garrido2009} and \cite{Consonni_2011_b}. More details about the specific growth conditions can be found elsewhere \cite{Fernandez-Garrido_nl_2013,Consonni_prb_2012,Fernandez-Garrido_nl_2012}.
The morphology of the samples was investigated by scanning electron microscopy (SEM). As shown in Fig.~\ref{SEM_Si}, the two samples grown on Si(111) (samples A and B) exhibit very different morphologies. Most NWs of sample A are significantly thicker than those of sample B, and frequently have irregular shapes suggesting NW coalescence. The coalescence degree of sample A thus appears to be higher than that of sample B. In contrast, the samples prepared on AlN/6H-SiC$(000\overline{1})$ (samples C and D) have a similar apparent coalescence degree which in both cases seems to be significantly higher than in the samples grown on Si(111) (see Fig.~\ref{SEM_SiC}).  

Low temperature (10 K) $\mu$PL was excited using the 325 nm line of a Kimon He-Cd laser, dispersed by an 80 cm Horiba Jobin Yvon monochromator and detected by a charge-coupled device detector. The spectral resolution of the system is approximately $0.25$~meV. 

XRD experiments were performed with Cu$K\alpha_{1}$ radiation (wavelength $\lambda=1.54056$~\AA) using a Panalytical X-Pert Pro MRD\texttrademark\ system equipped with a Ge$(220)$ hybrid monocromator. Symmetric $\theta/2\theta$ scans across the GaN $0002$, $0004$, and $0006$ Bragg reflections were measured with a three-bounce Ge$(220)$ analyzer crystal. The measurements were corrected for the resolution function which was obtained by measuring the same reflections from a bulk GaN$(0001)$ single crystal with a dislocation density lower than $10^{5}$~cm$^{-2}$ (purchased from AMMONO). A detailed description of the experimental determination of the resolution function can be found in Ref.\cite{Kaganer_prb_2012}. Both $\omega$ and $\varphi$ scans across the GaN $0002$ and $10\bar{1}5$ reflections, respectively, were recorded with a 1~mm slit in front of the detector instead of the analyzer crystal. $\varphi$ scans across the GaN $10\bar{1}2$ and $10\bar{1}1$ reflections were measured with an open detector.  

\section{Results}
\subsection{Quantitative evaluation of the coalescence degree}
In order to quantitatively evaluate the coalescence degree of the different samples, we employed the two methods proposed in Ref.\cite{Brandt_unpublished_2013}. Both methods are based on the analysis of the cross-sectional area and perimeter of the GaN NWs. To assess these parameters, we analyzed plan-view SEM images covering several hundreds of NWs by the open-source software ImageJ \cite{ImageJ}.
\begin{figure*}
\includegraphics*[width=\textwidth]{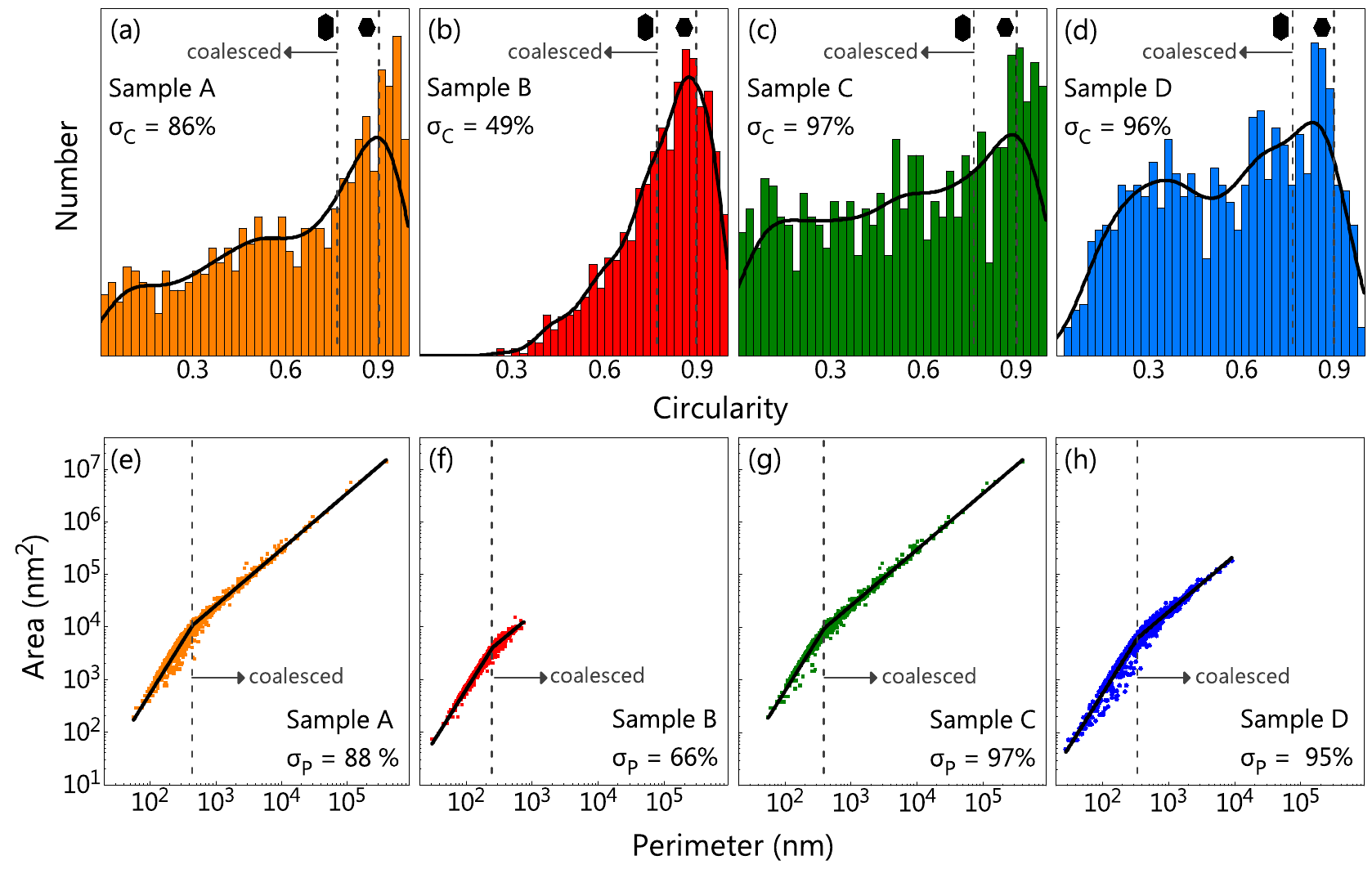}
\caption{\label{fig:coalescence degree} (Color online) Histograms of circularity and area-perimeter plots for the GaN NWs from samples A [(a) and (e)], sample B [(b) and (f)], sample C [(c) and (g)] and sample D [(d) and (h)], respectively. The solid lines in (a)–(d) show the kernel density estimations of the respective histograms. The vertical dashed lines indicate the circularity of the geometrical shape displayed next to them. The solid lines in (e)–(h) show fits of the data (solid symbols) by Eq.~(\ref{area vs perimeter}). The coalescence degrees obtained using the criterion of a minimum circularity of 0.762 (a--d) or the criterion of a critical perimeter $P_{C}$ (e--h) are also provided in each figure.}
\end{figure*}

The first method is based on the analysis of the circularity of the NW cross-sectional shapes. This parameter is defined as
\begin{equation}
\label{circularity}
C=4\pi{A}/P^{2},
\end{equation}
where $A$ and $P$ are the NW cross-sectional area and perimeter, respectively. It follows from this definition that $C=1$ for a circle, $C\approx0.907$ for a regular hexagon, and $C\ll1$ for a strongly elongated object. 

As discussed in detail in Ref.\cite{Brandt_unpublished_2013}, a circularity value of 0.762 is a conservative estimate for distinguishing single NWs and coalesced aggregates. Based on this criterion, the coalescence degree can be defined as
\begin{equation}
\label{sigma C}
\sigma_{C}=A_{C<0.762}/A_{T}.
\end{equation}
Here $A_{C<0.762}$ is the total cross-sectional area of those NWs with a circularity lower than 0.762, and $A_{T}$ is the total cross-sectional area of all NWs considered in the analysis. 

\begin{table}
\caption{\label{tab:1}Coalescence degrees $\sigma_{c}$ and $\sigma_{p}$, total linewidths $\Delta{E}$ of the (D$^{0}$,X$_{A}$) transition observed by $\mu$PL as well as tilt, twist and rms strain $\mathcal{E}$ of samples A--D.}
\resizebox{15cm}{!} {
\begin{tabular}{c c c c c c c}
\br
Sample&Substrate&$\sigma_{c}$/$\sigma_{p}$ (\%)&$\Delta{E}$ (meV)&tilt (deg)&twist (deg)&$\mathcal{E}$\\
\mr
A&Si$(111)$&86~/~88&$6.3\pm0.3$&1.6&2.0&$\left(3.7\pm0.2\right)\times10^{-4}$\\
B&Si$(111)$&49~/~66&$1.8\pm0.2$&3.8&3.7&$\left(1.5\pm0.2\right)\times10^{-4}$\\
C&AlN/6H-SiC$(000\overline{1})$&97~/~97&$3.0\pm0.1$&0.4&0.6&$\left(2.2\pm0.2\right)\times10^{-4}$\\
D&AlN/6H-SiC$(000\overline{1})$&96~/~95&$2.3\pm0.1$&0.4&0.6&$\left(2.0\pm0.2\right)\times10^{-4}$\\
\br
\end{tabular}
}
\end{table}

Figures~\ref{fig:coalescence degree}(a)--\ref{fig:coalescence degree}(d) present the histograms of the circularity for the different samples together with the coalescence degrees derived from Eq.~(\ref{sigma C}). Samples A and B, grown on Si$(111)$, exhibit a coalescence degree of 86\% and 49\%, respectively. In contrast, the coalescence degrees of the samples prepared on AlN/6H-SiC$(000\overline{1})$ are almost identical, namely, 97\% (sample C) and 96\% (sample D). However, the histograms of samples C and D reveal an interesting difference between these two samples. Unlike sample D, sample C exhibits a significant number of NWs with extremely low circularity. This is in good agreement with the visual inspection of Fig.~\ref{SEM_SiC}~(b), where many NWs seem to have coalesced into giant aggregates, the largest of which are close to span across the entire micrograph. Therefore, although the coalescence degrees of samples C and D are comparable, they are qualitatively different in that sample C is close to the percolation threshold.

The second method consists in the analysis of the area-perimeter plot dependencies shown in Figs.~\ref{fig:coalescence degree}(e)--\ref{fig:coalescence degree}(h). For all samples the slopes change abruptly from quadratic to nearly linear at a certain critical perimeter $P_{C}$. Those NWs with a perimeter larger than $P_{C}$ are considered to be not single NWs but coalesced aggregates \cite{Brandt_unpublished_2013}. Therefore, analogously to Eq.~(\ref{sigma C}), the coalescence degree is defined as
\begin{equation}
\label{sigma P}
\sigma_{P}=A_{P>P_{C}}/A_{T},
\end{equation}
where $A_{P>P_{C}}$ is the total cross-sectional area of those NWs with a perimeter larger than \textit{P$_{C}$}. The values of \textit{P$_{C}$} for the different samples were obtained by fitting the experimental data to the following function:
\begin{equation}
\label{area vs perimeter}
A=\alpha\left[P^{2}H\left(P_{C}-P\right)+P_{C}^{2-\beta}P^{\beta}H\left(P-P_{C}\right)\right],
\end{equation}
where \textit{P$_{C}$}, $\alpha$ and $\beta$ are fitting parameters, and \textit{H} is the Heaviside step function \cite{Brandt_unpublished_2013}.

As shown in Fig.~\ref{fig:coalescence degree} and summarized in Table~\ref{tab:1}, the coalescence degrees obtained by this second method agree very well with those derived from the analysis of the circularity, except for sample B. As explained in Ref.\cite{Brandt_unpublished_2013}, this discrepancy arises from the fact that the lower the coalescence degree, the less
data points constitute the linear portion of the area-perimeter plot and the fit to Eq.~(\ref{area vs perimeter}) tends to underestimate \textit{P$_{C}$}. Regardless of the method used to quantify the coalescence degree, it is clear that the coalescence degree is comparatively low for sample B, medium for sample A, and high for samples C and D. 

\subsection{Low temperature photoluminescence}
\begin{figure}
\hspace{1in}\includegraphics {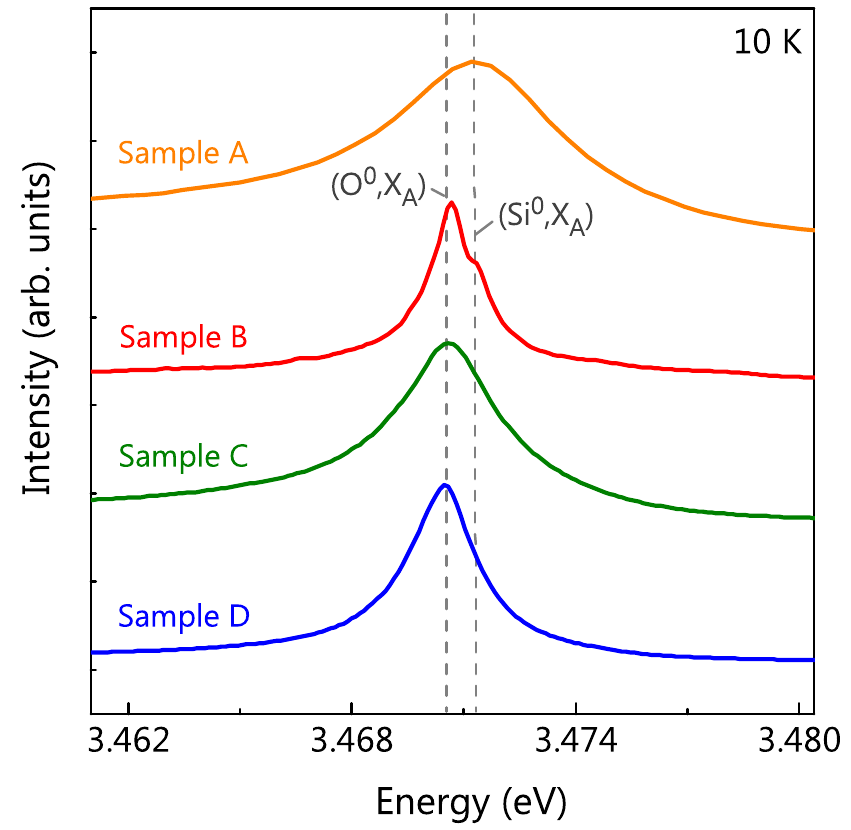}
\caption{\label{fig:LTPL}(Color online) Near band-edge low-temperature $\mu$PL spectra of samples A--D. The dashed lines indicate the position of the transitions related to A excitons bound to neutral Si [(Si$^{0}$,X$_{\rm{A}}$)] and O donors [(O$^{0}$,X$_{\rm{A}}$)] in the GaN NWs.}
\end{figure}

Figure~\ref{fig:LTPL} shows the near band-edge low temperature $\mu$PL spectra of the GaN NW samples under investigation. As commonly observed for spontaneously formed GaN NWs, the PL spectra are dominated by the recombination of A excitons bound to neutral Si and O donors (D$^{0}$,X$_{\rm{A}}$) at approximately 3.471 eV. The two different transitions can be distinguished in sample B. In agreement with previous works reported in the literature, the position of the (D$^{0}$,X$_{\rm{A}}$) line fits with the expected one for a fully relaxed GaN crystal \cite{Calleja2000,Brandt_prb_2010}. The total linewidth $\Delta{E}$ varies between 1.8 and 6.3~meV and does not exhibit a monotonic increase with the coalescence degree (see Table~\ref{tab:1}). 

\subsection{Structural properties}
\begin{figure}
\hspace{1in}\includegraphics {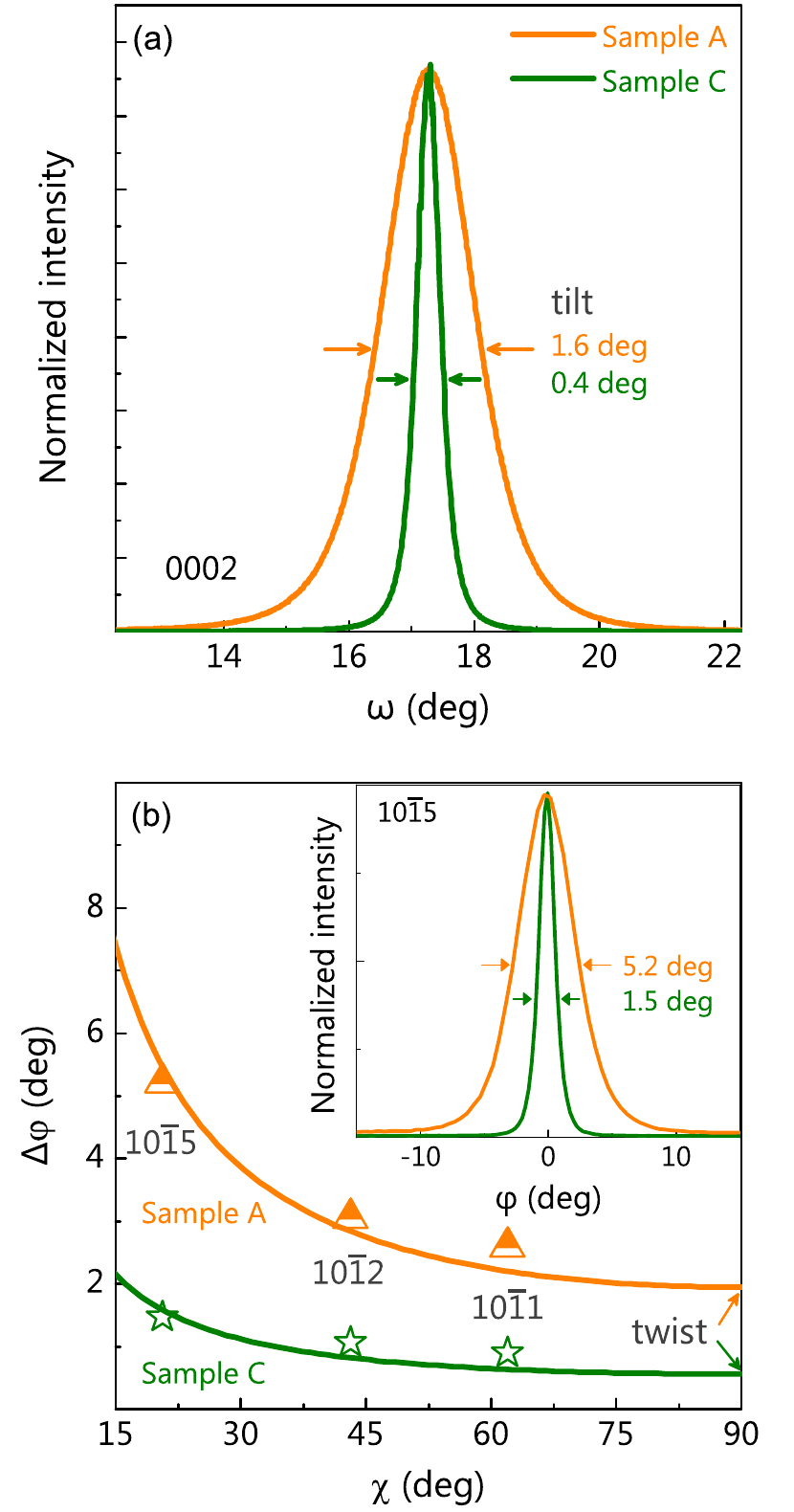} 
\caption{\label{fig:XRDa}(Color online) (a) XRD $\omega$ scans across the symmetric GaN $0002$ reflection for samples A and C. (b) FWHM (solid circles) of the maxima of XRD $\varphi$ scans across the GaN $10\bar{1}5$, $10\bar{1}2$ and $10\bar{1}1$ reflections in skew geometry as a function of the tilt angle $\chi$ for samples A and C. The solid lines display the fits of the data by Eq.~(\ref{equTwist}). The inset shows the $\varphi$ scans across the GaN $10\bar{1}5$ reflection.}
\end{figure}

Although the strain induced by NW coalescence should broaden the excitonic transitions observed in the $\mu$PL spectra of the GaN NWs, the results shown above seem to indicate that there is no direct correlation between the coalescence degree and the total linewidth of the donor bond exciton transition. However, the strain induced by NW coalescence should depend not only on the coalescence degree but also on the mutual misorientation of adjacent NWs. In the following, we investigate both the orientation distribution of the NW ensembles and their strain state by XRD.

The out-of-plane orientation distribution, i.\,e., the tilt, is assessed by $\omega$ scans across the GaN $0002$ reflection. As shown exemplarily in Fig.~\ref{fig:XRDa}~(a) for samples A and C and summarized in Table~\ref{tab:1}, the NWs prepared on Si$(111)$ exhibit a comparatively broad out-of-plane orientation distribution. The tilt values of 1.6$^{\circ}$ and 3.8$^{\circ}$ measured for samples A and B are typical for GaN NWs grown on bare Si$(111)$ \cite{Geelhaar_ieeejstqe_2011,Jenichen2011a,Wierzbicka_natech_2012}. The lower value for the tilt measured in sample A is a direct consequence of its higher degree of coalescence. This decrease occurs because the tilt of highly misoriented NWs is partially accommodated upon NW coalescence. The tilt is significantly reduced to a value of 0.4$^{\circ}$ for the two samples grown on AlN/6H-SiC$(000\bar{1})$. This narrow out-of-plane orientation distribution is the result of the well-defined epitaxial relation between the AlN buffer layer and the GaN NWs \cite{Fernandez-Garrido_nl_2012}.

\begin{figure}
\hspace{1in}\includegraphics{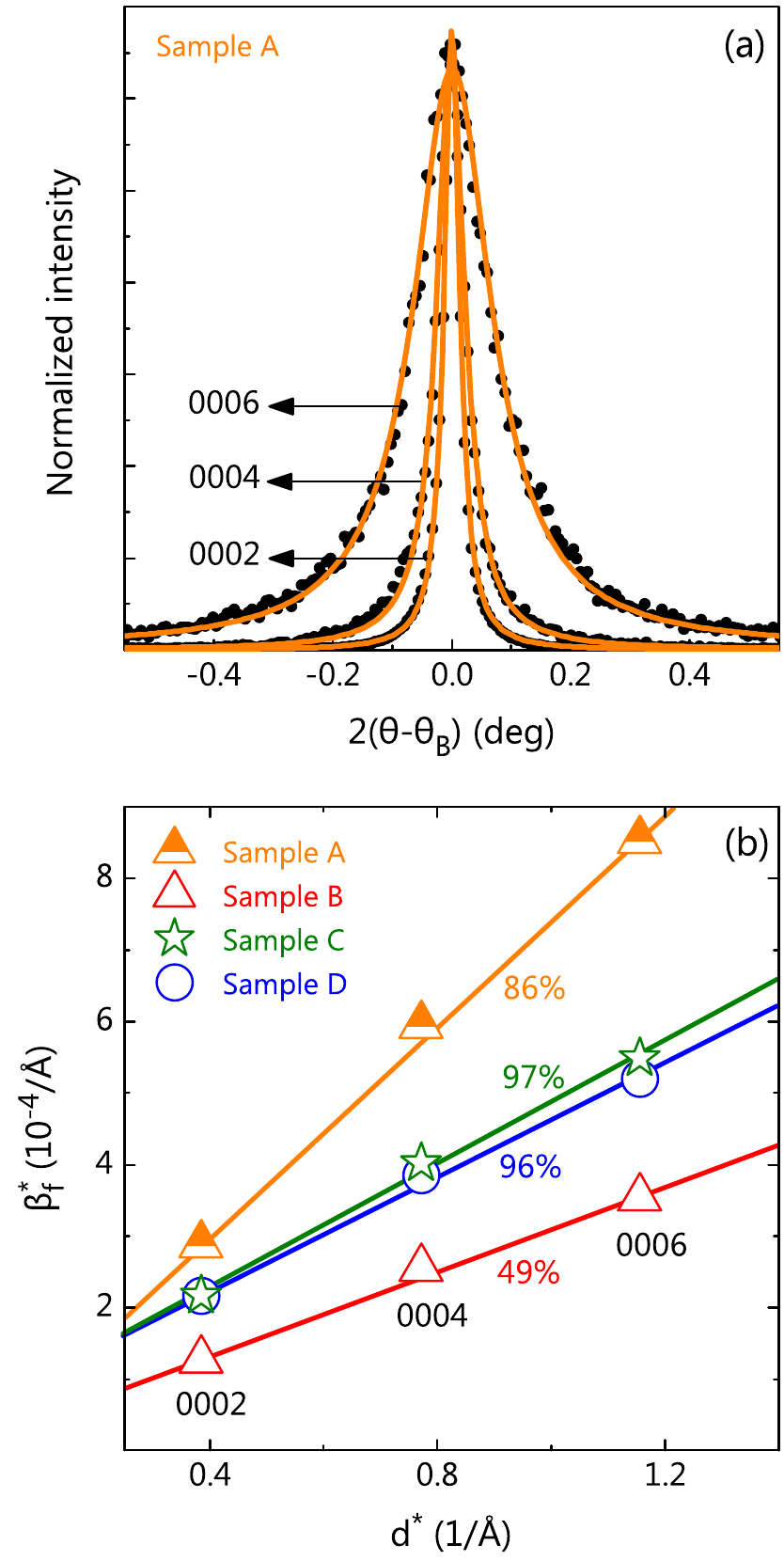}
\caption{\label{fig:XRD}(Color online) (a) XRD $\theta/2\theta$ scans across the $00.n$ reflections with $n=2,4,6$ for sample A. The solid lines represent fits of the data by a Lorentzian. (b) Williamson-Hall plot for the symmetric GaN reflections of samples A--D. The solid lines are fits of the data by Eq.~(\ref{equXRD}). The coalescence degrees obtained using the criterion of a minimum circularity of 0.762 are indicated for comparison.}
\end{figure}

To assess the in-plane orientation distribution, i.\,e., the twist of the NW ensembles, we recorded $\varphi$ scans across the GaN $10\bar{1}5$, $10\bar{1}2$ and $10\bar{1}1$ reflections in skew geometry. Their full widths at half maxima (FWHMs) $\Delta\varphi$ are plotted in Fig.~\ref{fig:XRDa}~(b) as a function of the tilt angle $\chi$. According to Ref.~\cite{Jenichen2011a}, the variation of $\Delta\varphi$ with $\chi$ can be described by the following equation: 
\begin{equation}
\label{equTwist}
\Delta\varphi(\chi)=\Delta\varphi(90^{\circ})/[\Delta\varphi(90^{\circ})/360^{\circ}+\sin(\chi)],
\end{equation}
where $\Delta\varphi(90^{\circ})$ is the FWHM for the GaN $10\bar{1}0$ reflection, i.\,e., the true twist of the NWs \cite{Jenichen2011a}. Fits of Eq.~(\ref{equTwist}) to the experimental data yield a twist of 0.6$^{\circ}$ for the samples grown on AlN/6H-SiC$(000\overline{1})$. Similar to the tilt, this value is much lower than those measured for the NWs grown on bare Si$(111)$, namely, 2.0 and 3.7$^{\circ}$ for samples A and B, respectively. The NWs prepared on AlN/6H-SiC$(000\bar{1})$ thus also exhibit a significantly narrower in-plane orientation distribution. 

To determine the strain state of the samples, XRD $\theta/2\theta$ scans were carried out. The positions of the GaN $000n$ out-of-plane reflections are the same for all samples, indicating that, independently of the substrate, all NW ensembles are, on average, free of homogeneous strain, i.\,e., $\left\langle\varepsilon_{zz}\right\rangle=0$ where $\varepsilon_{zz}$ is the $zz$-component of the strain tensor. This result is consistent with those obtained by PL spectroscopy. However, as shown in Fig.~\ref{fig:XRD}~(a) for sample A, a comparison of the $\theta/2\theta$ scans reveals that the linewidths of the successive reflections steadily increase with the reflection order $n$. The same trend is observed in the successive reflection orders of samples B, C, and D (not shown).

The diffraction peak broadening with the reflection order is an indication of the presence of inhomogeneous strain. It can be characterized by the rms strain, also referred to as the micro-strain, defined as   
\begin{equation}
\label{equStrain}
\mathcal{E}=\left\langle\varepsilon^{2}_{zz}\right\rangle ^{1/2}.
\end{equation}
Since our analysis is restricted to the symmetric Bragg reflections, only the $\varepsilon_{zz}$ strain component is involved.

To actually obtain $\mathcal{E}$ for the different samples, we employ Williamson-Hall plots \cite{Jenichen2011a,Langford_2000} in the reciprocal space representation given by $\beta_{f}^{*}=\beta_{f}'\cos\left(\theta\right)/\lambda$ and $d^{*}=2\sin\left(\theta\right)/\lambda$ [Fig.~\ref{fig:XRD}~(b)]. Here, $\beta_{f}'$ is the integral breadth of the diffraction profile corrected by the breadth of the resolution function. Lorentzian line shapes were used to fit all peaks. The value of $\mathcal{E}$ can be obtained from the slope of the linear fit to the Williamson-Hall plot, using the relation
\begin{equation}
\label{equXRD}
\beta_{f}^{*}=\beta_{s}^{*}+2\mathcal{E}d^{*},
\end{equation}
where $\beta_{s}^{*}$ represents the broadening of the line due to size effects. The resulting values of $\mathcal{E}$ are summarized in Table~\ref{tab:1}. For the samples grown on Si(111), we found that $\mathcal{E}$ increases significantly with the coalescence degree, from $1.5\times10^{-4}$ (sample B) to $3.7\times10^{-4}$ (sample A). Despite the even larger coalescence degree of samples C and D, smaller values of $\mathcal{E}$ were measured, namely, $2.0\times10^{-4}$ (sample D) and $2.2\times10^{-4}$ (sample C).  

\section{Discussion}

If the lattice disorder at the interface between the substrate and the GaN NWs were the dominant source of inhomogeneous strain in NW ensembles, we would expect the samples prepared on bare Si$(111)$ to always exhibit either higher or lower values of $\mathcal{E}$ compared to those on AlN/6H-SiC$(000\overline{1})$. Experimentally, however, we observe the contrary: the two samples on Si$(111)$ exhibit the lowest and highest value of $\mathcal{E}$ of all samples. In principle, a larger contribution to $\mathcal{E}$ for the NWs prepared on Si$(111)$ is expected since the NWs on AlN/6H-SiC$(000\overline{1})$ exhibit a well-defined epitaxial interface to the substrate \cite{Fernandez-Garrido_nl_2012}, where the lattice mismatch is accommodated by means of misfit dislocations. While misfit dislocations also induce inhomogeneous strain, their impact is expected to be not as strong as that of the locally broken and distorted bonds at the interface between the NWs and the amorphous Si$_{x}$N$_{y}$ interlayer on Si$(111)$. The fact that the samples grown on Si$(111)$ exhibit both the lowest and the highest values of $\mathcal{E}$ clearly shows that the main source of inhomogeneous strain in GaN NW ensembles is NW coalescence.

The non-monotonic increase of $\mathcal{E}$ with the coalescence degree observed in this work can then be easily explained: the magnitude of $\mathcal{E}$ originating from NW coalescence depends on the mutual misorientation of adjacent NWs. This additional factor explains why the values of $\mathcal{E}$ for samples C and D are almost a factor of two lower than for sample A despite their very high coalescence degree. Therefore, apart from reducing the coalescence degree, an efficient and feasible route to minimize the impact of NW coalescence consists in decreasing the tilt and twist of the NWs by growing them epitaxially on suitable substrates. 

\begin{figure}
\hspace{1in}\includegraphics{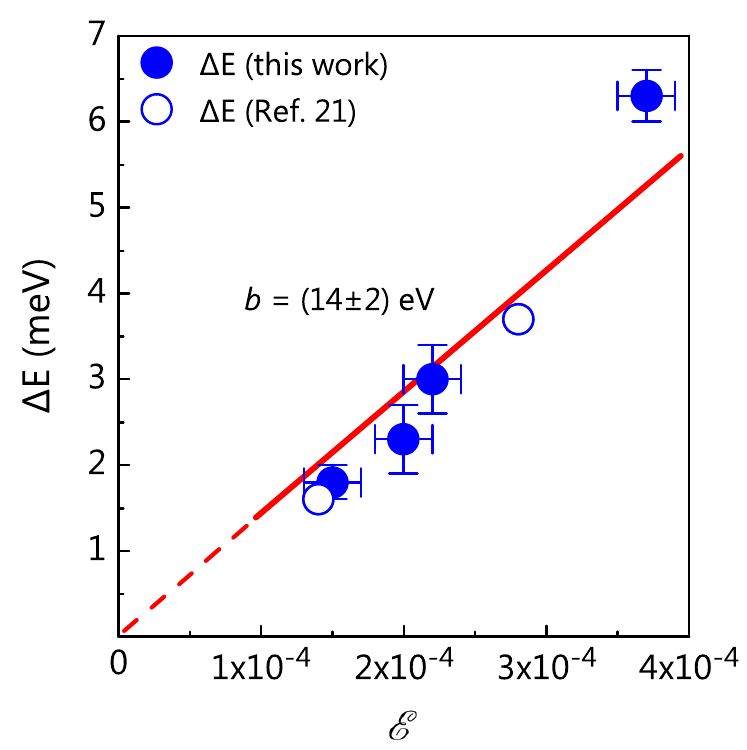}
\caption{\label{fig:summary}(Color online) Linewidths of the (D$^{0}$,X$_{\rm{A}}$) transition in the $\mu$PL spectra of samples A--D as a function of $\mathcal{E}$. The values reported in Ref.\cite{Jenichen2011a} for GaN NWs grown on bare Si$(111)$ are also included. The solid line is a linear fit.}
\end{figure}

Strain in NWs should cause not only the broadening of XRD profiles but should also influence the linewidth of PL transitions. 
For $\varepsilon_{zz}<10^{-4}$, the energy of an excitonic transition changes linearly with this strain component. Since the strain inhomogeneities detected in samples A--D are of the same order of magnitude, we expect a linear dependence between the rms strain assessed by XRD and the FWHM $\Delta E$ of the PL transitions. Figure~\ref{fig:summary} presents the linewidth of the (D$^{0}$,X$_{\rm{A}}$) transition as a function of $\mathcal{E}$ for samples A--D. We also include the experimental data reported in Ref.~\cite{Jenichen2011a} for GaN NW ensembles grown on bare Si$(111)$. Evidently, $\Delta E$ increases monotonically with $\mathcal{E}$. A linear fit of the data yields a slope $b=(14\pm 2)$~eV. This value should be compared with the deformation potentials that describe the variation in the bandgap with the different strain components \cite{Ishii_prb_2010}. This comparison, however, is not straightforward because the coalescence of NWs presumably generates a rather complex strain field and our XRD experiments only reveal one strain component, $\epsilon_{zz}$. In any case, the slope we observe is close to the value of $16.5$~eV expected for pure biaxial strain \cite{Ghosh_prb_2002}.

We note that, besides the inhomogeneous strain, the position of the (D$^{0}$,X$_{\rm{A}}$) transition may also vary as much as a few meV because the energy of the donor bound-exciton states depends on their distance to the NW sidewall surfaces \cite{Brandt_prb_2010,Corfdir2009,Corfdir_jap_2012}. Therefore, for sufficiently low values of $\mathcal{E}$, we expect a deviation from the linear behavior shown in Fig.~\ref{fig:summary} and a linewidth still greatly exceeding that of a perfect GaN bulk crystal ($\ll 0.1$~meV). Given that the impact of surface induced effects on the broadening of the (D$^{0}$,X$_{\rm{A}}$) transition depends on the
specific cross-sectional shape and diameter of the NWs, this phenomenon might also be partly responsible for the scatter of the experimental data. 

\section{Conclusions}
The quantitative evaluation of the coalescence degree in GaN NW ensembles grown on different types of substrates has allowed us to analyze the impact of NW coalescence on their structural and optical properties. The main conclusions of the present work are (i) the inhomogeneous strain detected by XRD in NW ensembles is mainly caused by the coalescence of closely spaced NWs, (ii) the magnitude of the coalescence-induced strain inhomogeneity depends on both the coalescence degree and the mutual misorientation of adjacent NWs, and (iii) the linewidth of the excitonic transitions observed by PL spectroscopy does not exhibit a monotonic increase with the coalescence degree but scales with the rms strain.

The rms strain observed in this work is an average value characteristic of the entire NW ensemble. Therefore, this quantity could vary substantially from NW to NW and/or within individual NWs. In fact, it is even possible that, although the NW ensemble exhibits a net strain of zero on average, the individual NWs are alternately under compressive and tensile strain such that $\left\langle\varepsilon_{zz}\right\rangle=0$. In this context, it would be highly interesting to analyze the strain state of single NWs using x-ray nanodiffraction \cite{Gulden_pssa_2011,Biermanns_jac_2012} and compare the results with those obtained for the corresponding NW ensemble. Furthermore, the correlation of the strain state of single NWs with their PL spectra would provide valuable information to understand the physical mechanisms governing the linewidth of the excitonic transitions observed in GaN NWs.

\ack 

We would like to thank Anne-Kathrin Bluhm for providing the scanning electron micrographs presented in this work, Hans-Peter Sch\"{o}nherr for his dedicated maintenance of the MBE system, and Pinar Dogan for a critical reading of the manuscript. Finally, we are indebted to Henning Riechert for continuous encouragement and support.


\bibliographystyle{iopart-num}
\bibliography{bibliography}

\end{document}